\documentclass{article}

\usepackage{latexsym}
\usepackage{amssymb}
\usepackage{epsfig}
\usepackage{amsmath}
\usepackage{amsthm}
\usepackage{amstext}
\usepackage{amscd}

\def\ket#1{\left|#1\right>}

\newtheorem{defi}{Definition}
\newtheorem{theo}{Theorem}
\newtheorem{lem}[theo]{Lemma}
\newtheorem{cor}[theo]{Corollary}

\newtheorem{con}{Conjecture}
\newtheorem{prop}[theo]{Proposition}

\def\D{{D}}

\def\ket#1{|\,#1\,\rangle}

\newcommand{\textcirc}{% circle of reasonable size 
  \setlength{\unitlength}{1.5ex}%
  \begin{picture}(1.1,1)(-.55,-.55)%
    \put(0,0){\circle{1.1}}%
  \end{picture}}
\renewcommand{\bar}{\overline}

\newcommand{\defend}{\hspace*{\fill} $\textcirc$}
\newcommand{\exend}{\hspace*{\fill} $\diamondsuit$}
\newcommand{\Tr}{{\rm Tr}}
\newcommand{\HA}{{\cal H}_{{A}}}
\newcommand{\HB}{{\cal H}_{{B}}}
\newcommand{\HE}{{\cal H}_{{E}}}
\newcommand{\oz}{\overline{z}}
\newcommand{\prob}{{\rm Prob\,}}
\newcommand{\cd}{{D}}

\newcommand{\noi}{\noindent}
\newcommand{\cancel}[1]{}
\newcommand{\canpro}[1]{}

\newcommand{\be}{\beta}
\newcommand{\ee}{\end{equation}}
\newcommand{\bea}{\begin{eqnarray}}
\newcommand{\beq}{\begin{equation}}
\newcommand{\eea}{\end{eqnarray}}
\newcommand{\beas}{\begin{eqnarray*}}
\newcommand{\eeas}{\end{eqnarray*}}
\newcommand{\bece}{\begin{center}}
\newcommand{\ence}{\end{center}}
\newcommand{\beit}{\begin{itemize}}
\newcommand{\enit}{\end{itemize}}

\newcommand{\pe}{\hspace*{\fill} $\Box$\\ \ \\}
\newcommand{\peon}{\hspace*{\fill} $\Box$\\}

\newcommand{\z}{{\bf Z}}

\newcommand{\al}{\alpha}

\newcommand{\ep}{\varepsilon}
\newcommand{\de}{\delta}

\newcommand{\ga}{\gamma}

\newcommand{\op}{\oplus}

\newcommand{\OX}{\overline{X}}

\newcommand{\CE}{{{\cal E}}}

\newcommand{\OY}{\overline{Y}}
\newcommand{\OZ}{\overline{Z}}

\newcommand{\cx}{{\cal X}}

\newcommand{\cy}{{\cal Y}}

\newcommand{\CX}{{\cal X}}
\newcommand{\CY}{{\cal Y}}
\newcommand{\CZ}{{\cal Z}}

\newcommand{\gd}{\Delta}

\newcommand{\ida}{I(X;Y\hspace{-1.3mm}\downarrow\hspace{-0.8mm} Z)}

 \def\<{\langle}
 \def\>{\rangle}
 
 \def\down{\downarrow}

 \def\opone{\leavevmode\hbox{\small1\kern-3.8pt\normalsize1}}
 \def\H{{\cal H}}

 \def\z{{\bar z}}
 \newcommand{\complex}{{\kern .1em {\raise .47ex\hbox {$\scriptscriptstyle
 |$}}\kern -.4em {\rm C}}}
 \newcommand{\real}{{{\rm I} \kern -.19em {\rm R}}}
\newcommand{\eeq}{\end{equation}}
 \newcommand{\beqa}{\begin{eqnarray}}
 \newcommand{\eeqa}{\end{eqnarray}}

\begin{document}

\title{\bf The Impossibility of Pseudo-Telepathy\\ Without Quantum 
Entanglement}

\author{Viktor Galliard\, \footnote{Untervaz, Switzerland. E-mail: math@galliard.ch.
 Supported by Canada's NSERC and Swisscom IT Services Ltd.}
\qquad
Alain Tapp\, \footnote{D\'epartement 
d'Informatique et recherche op\'erationnelle,
Universit\'e de Montr\'eal,
C.P. 6128 succ Centre-Ville,
Montr\'eal, Qu\'ebec,  H3C 3J7,
Canada. E-mail:
tappa@iro.umontreal.ca}
\qquad Stefan Wolf\, \footnote{D\'epartement 
d'Informatique et recherche op\'erationnelle,
Universit\'e de Montr\'eal,
C.P. 6128 succ Centre-Ville,
Montr\'eal, Qu\'ebec,  H3C 3J7,
Canada. E-mail:
wolf@iro.umontreal.ca}}

\date{}

\maketitle

\begin{abstract}
\noi
Imagine that  Alice and Bob, unable to communicate, are both given 
a $16$-bit string such that the strings are either equal, or they
 differ in exactly 
$8$ positions. Both parties are then supposed to output a $4$-bit string 
in such a way that these short strings 
are equal if and only if the original longer strings given to them were equal as well. It is
known that this task can be fulfilled without failure and without communication
if Alice and Bob share $4$ maximally entangled quantum bits. We show that,
on the other hand, they {\em cannot\/} win the same game with certainty
if they only share {\em classical\/} bits, even if it is an unlimited number.
This means that for fulfilling this particular distributed task, quantum entanglement 
can completely replace communication. This phenomenon has been called pseudo-telepathy.
The results of this paper complete the analysis of the first proposed game of this type 
  between two players.
\end{abstract}

%\newpage

\section{Introduction}

\subsection{Pseudo-Telepathy Games}

{\em Pseudo-telepathy\/}  is a phenomenon showing that for achieving certain 
well-defined distributed tasks,  communication can be replaced by measuring shared
quantum states proving so-called {\em entanglement}; this does {\em not\/} imply,
however, that such quantum entanglement allows for, instantaneous, communication.

More specifically, a {\em two-player pseudo-telepathy game\/} is a game in which two 
separated parties 
who are not able to communicate 
are asked two {\em questions}, $x_A$ and $x_B$, respectively, and should give 
{\em answers\/} $y_A$ and $y_B$ satisfying a certain condition defined by the 
game. Formally, given that the pair of questions $(x_A,x_B)$ belongs to 
a certain relation $R_X\subseteq {\cal X}_A\times {\cal X}_B$, the answers 
have to be such that 
\[
(x_A,x_B,y_A,y_B)\in R_{XY}\subseteq {\cal X}_A\times {\cal X}_B\times {\cal Y}_A\times {\cal Y}_B
\]
holds, where the game is defined by the relations $R_X$ and $R_{XY}$. 
(Here, ${\cal X}_A$, ${\cal X}_B$, ${\cal Y}_A$, and ${\cal Y}_B$ 
stand for the ranges of possible questions to and answers from Alice 
and Bob, respectively.) 

Some of these games are of particular interest since  they can be won
by parties sharing quantum information, but {\em not\/} by parties sharing only 
classical information initially. A game with this property can be used for demonstrating 
the existence of quantum entanglement|however, this is true only if a proof is provided
that there is no classical strategy for winning the game with certainty. The main 
result of this paper is to provide this proof for the first and most prominent example 
of such a game~\cite{BCT99}, for which, previously, only an asymptotic impossibility 
proof has been given  
(i.e., no concrete parameters have been known for which 
the game cannot be won).

\subsection{The Game by Brassard, Cleve, and Tapp}
 
In~\cite{BCT99}, the following two-party game was proposed. Let $n\geq 1$, and let $N=2^n$. 
In the game with parameter $n$, the questions asked to the parties, Alice and 
Bob, are arbitrary $N$-bit strings $x_A$ and $x_B$ satisfying 
\beq\label{dj}
d_H(x_A,x_B)\in \{0,N/2\}
\ee
(i.e., the questions are either equal or differ in exactly half the positions). 
Alice's and Bob's answers $y_A$ and $y_B$ have to be binary strings of length 
$n=\log_2 N$ satisfying the simple condition 
\[
y_A=y_B\ \longleftrightarrow\  x_A=x_B\ .
\]

It was shown in~\cite{BCT99} that this game can always be won if Alice and Bob
share $n$ maximally entangled quantum bits (so-called EPR pairs). On the other 
hand, a proof was also given that the game cannot be won without quantum 
entanglement 
if $n$ is large enough. Unfortunately, this purely asymptotic result is not satisfactory
since $n$ has to take a particular, fixed, value, should the game be 
actually played (e.g., performed as an experiment demonstrating quantum entanglement).
Moreover, the game {\em can\/} be won classically {\em with high probability\/} for large $n$
if Alice and Bob agree to respond their respective questions with a random 
$n$-bit hash value thereof (using the same, predetermined, hash function).

Motivated by this, we address the question which  the smallest value of $n$ is for 
which the game cannot be won.
In fact, for the choices $n=1,2$, and $3$, the game can be won classically with probability 
$1$~\cite{BCT99},\cite{gw02}. It was conjectured that for $n=4$ this is not the case 
anymore~\cite{gw02}. In the present paper, we prove that this conjecture is true.
We will  use of a connection that has been made in~\cite{gw02} between the game 
and graph colorings. In fact, the main part of the proof will be to derive a lower
bound on the chromatic number of a certain graph $G$.

\section{A Classical Impossibility Result}

In this section we prove the following main result of the paper.

\begin{theo}\label{hr}
Classically, the described pseudo-telepathy game cannot be won with certainly  
for $n=4$.
\end{theo}

\noi
In~\cite{gw02}, the following connection was made between the pseudo-telepathy 
game and graph colorings. Let $G_N=(V_N,E_N)$ be the graph defined by 
\begin{eqnarray*}
V_N & = & \{0,1\}^N\\
E_N  & = & \{(u,v)\in V_N^2\, |\, d_H(u,v)=N/2\}\ .
\end{eqnarray*}

\noi
Then the game with parameter $N=2^n$ can be won classically if and only of 
\beq\label{gi}
\chi(G_N)\leq N\ ,
\ee
where $\chi(G_N)$ is the chromatic number of $G_N$, i.e., the minimum number of 
different colors necessary to color the vertices of $G_N$ in such a way 
that no vertices with the same color are connected by an edge. If inequality
(\ref{gi}) is satisfied, then 
Alice and Bob's strategy for winning the game is to agree on a coloring of the 
graph beforehand and to answer a question, i.e., a vertex of the graph, by its 
color. 
If on the other hand (\ref{gi}) is violated, then the game cannot be won 
with certainty: A winning strategy {\em is\/} a corresponding 
coloring.
All we have to prove therefore in order to obtain Theorem~\ref{hr}
is the following.

\begin{prop}\label{hp}
Let $G=G_{16}$ be the graph as defined above. Then 
\beq\label{cu}
\chi(G)>16\ .
\ee
\end{prop}

\noi
In order to prove Proposition~\ref{hp}, we use that fact that 
\beq\label{gg}
\chi(G)\geq \frac{|V(G)|}{M}
\ee
holds
if $M$ is an upper bound on the size of all independent sets of the graph. An independent 
set 
is a set of vertices which are pairwisely unconnected, and clearly, any set of vertices
of the same color in a coloring is independent.

Because of  (\ref{hp}) and (\ref{gg}), it is sufficient to show that no independent set 
of $G$ can be larger than 
\[
\frac{|V(G)|}{16}-1=4095\ .
\]

\begin{lem}\label{hl}
Let $I\subseteq V$ be an independent set of $G$. Then 
\[
|I|\leq 3912\ .
\]
\end{lem}

\proof
Let us simplify the problem as follows. First, we observe that the graph $G$
consists of two isomorphic connected components $G_e$ and $G_o$ (containing 
the vertices of even and odd Hamming weight, respectively). Secondly, 
a maximum independent set contains a vertex $v$ if and only if it also 
contains its bitwise complement $\overline{v}$ since for all vertices $w$,
we have 
\[
d_H(v,w)=8\ \longleftrightarrow\  d_H(\overline{v},w)=8\ .
\]

Let $G_{e,<8}$ be the subgraph of $G_e$ containing the vertices of Hamming
weight less than $8$, let, for $i=0,2,4$, and $6$ $G_i$ be the subgraph 
of $G_{e,<8}$ containing the vertices of Hamming weight $i$, and let for 
a graph $H$ $M(H)$ denote the size of a maximum independent set of $H$. 
Since 
\begin{eqnarray*}
M(G) & = & 2M(G_e)\ \ =\ \ 4M(G_{e,<8})\\
 & \leq & 4(M(G_0)+M(G_2)+M(G_4)+M(G_6))\ ,
\end{eqnarray*}
it is sufficient to prove that 
\beq\label{ebe}
M(G_0)+M(G_2)+M(G_4)+M(G_6)\leq 3912/4=978
\ee
holds. We will show 
\begin{eqnarray}
M(G_0) & = & 1 \label{eis}\\
M(G_2) & = & 120 \label{zwo}\\
M(G_4) & = & 455 \label{drue}\\
M(G_6) & \leq & 402 \label{sibze} 
\end{eqnarray}
which implies (\ref{ebe}).
\ \\ \

\noi
{\it Proof of (\ref{eis}).}
Trivial.
\ \\ \

\noi
{\it Proof of (\ref{zwo}).}
We have \[M(G_2)=|G_2|={18 \choose 2}=120\] since none of the  vertices are connected. 
\ \\ \

\noi
{\it Proof of (\ref{drue}).}
The set of all vertices with a $1$ in the first position is independent, and it has
size $455$. Its maximality follows from a result by 
Erd\"os {\it et.\ al.}~\cite{EKR61},\cite{Wi84}. 
Let $m(n,k,t)$ be the maximum size of a subset $X$ of all $k$-element
subsets of $\{1,2,\ldots,n\}$ such that for all $a,b\in X$, we have 
$|a\cup b|\geq t$. The mentioned result states
\[
m(n,k,t)\leq {n-t \choose k-1}
\]
whenever $n\geq (k-t+1)(t+1)$.
We have $M(G_4)=m(16,4,1)$, and (\ref{drue}) follows.
\ \\ \

\noi
{\it Proof of (\ref{sibze}).}
Note first that for any independent set $I$ of $G_6$ such that any two vertices 
$u,v\in I$ satisfy $d_H(u,v)\leq 6$, we have 
\[
|I|\leq m(16,6,3)\leq 286
\]
according to the result used in the proof of (\ref{drue}).

Let us now  deal with the cases of independent sets $I$ containing $u,v\in I$ with 
$d_H(u,v)\in \{10,12\}$. 

Let first $u,v\in I$ with $d_H(u,v)=12$. Without loss of generality, we can assume 
\[ u=111111 \cdot 000000 \cdot 0000 \ , \ \ v=000000 \cdot 111111 \cdot 0000\ . \]
A third vertex $w\in I\setminus\{u,v\}$ has to be one of the following, 
modulo permutation of bit positions:
\[
\begin{array}{ccc}
000000 \cdot 111000 \cdot 1110 &
000000 \cdot 111100 \cdot 1100 &
000000 \cdot 111110 \cdot 1000 \\
100000 \cdot 100000 \cdot 1111 &
100000 \cdot 111000 \cdot 1100 &
100000 \cdot 111100 \cdot 1000 \\
100000 \cdot 111110 \cdot 0000 &
111000 \cdot 111000 \cdot 0000 &
\\
\end{array}
\]
Let now, for fixed $u$, $v$, and $w$, $G_{6,uvw}$ be the subgraph of $G_6$ arising 
when $u$, $v$, $w$, and their neighbors are removed. In order to find an upper 
bound on $M(G_{6,uvw})$, we decompose the graph into disjoint cliques, the number 
of which is such a bound. Using a simple greedy algorithm, we found decompositions 
for all possible vertices $w$ (or, more precisely, ``types'' of vertices as listed 
above). The following table shows for all of the $8$ types $i=1,\ldots,8$ the 
number $a_i$  of vertices of the type and the number $b_i$ of cliques in the decomposition 
of the corresponding graph.
\ \\
\begin{center}
\begin{tabular}{c|ccccccccc}
$i$         & 1   & 2   & 3   & 4   & 5   & 6   & 7   & 8   \\
\hline 
$a_i$     & 80  & 90  & 24  & 36  & 720 & 360 & 36  & 400 \\
$b_i$ & 394 & 426 & 495 & 320 & 370 & 399 & 425 & 314
\end{tabular}
\end{center}
\ \\
Since 
\[
\max\left\{\sum_{i\in \{1,2,3,4,7\}}{a_i}\ ,\ \max_{i\in \{5,6,8\}}\{b_i\}\right\}=399\ ,
\]
we get the bound $402=399+3$ on the size of an independent set of $G_6$ (including 
also $u$, $v$, and $w$). 
In order to see this, note that any independent set of $G_{6,uvw}$ of size
larger than $399$ would necessarily contain at least one vertex of one 
of the types $5$, $6$, or $8$, thus the bound obtained for these
types apply.
The relevant clique decompositions of $G_{6,uvw}$ if $w$ 
is of  the types $5$, $6$, and $8$ can be found at~\cite{web}.

The case of $u,v\in I$ with $d_H(u,v)=10$ can be treated similarly.
For 
\[ u = 11111 \cdot 1 \cdot 00000 \cdot 00000\ , \ \ v = 00000 \cdot 1 \cdot 11111 \cdot 00000\ , \]
 the list of types is 
\[
\begin{array}{ccc}
00000 \cdot 0 \cdot 10000 \cdot 11111 &
00000 \cdot 0 \cdot 11100 \cdot 11100 &
00000 \cdot 0 \cdot 11110 \cdot 11000 \\
00000 \cdot 0 \cdot 11111 \cdot 10000 &
00000 \cdot 1 \cdot 00000 \cdot 11111 &
00000 \cdot 1 \cdot 11000 \cdot 11100 \\
00000 \cdot 1 \cdot 11100 \cdot 11000 &
00000 \cdot 1 \cdot 11110 \cdot 10000 &
10000 \cdot 0 \cdot 10000 \cdot 11110 \\
10000 \cdot 0 \cdot 11100 \cdot 11000 &
10000 \cdot 0 \cdot 11110 \cdot 10000 &
10000 \cdot 0 \cdot 11111 \cdot 00000 \\
11000 \cdot 1 \cdot 11000 \cdot 10000 &
11000 \cdot 1 \cdot 11100 \cdot 00000 &
11100 \cdot 0 \cdot 11100 \cdot 00000 
\end{array}
\]
The clique decompositions found are of size
\\ \
\begin{center}
\begin{tabular}{c|ccccccccc}
$i$         & 1   & 2   & 3   & 4   & 5   & 6   & 7   & 8    \\
\hline 
$a_i$     & 5   & 100 & 50  & 5   & 1   & 100 & 100 & 25  \\
$b_i$ & 318 & 366 & 394 & 428 & 260 & 345 & 365 & 408
\end{tabular}

\begin{tabular}{c|ccccccccc}
$i$         & 9   & 10  & 11  & 12  & 13  & 14  & 15   \\
\hline 
$a_i$     & 125 & 500 & 125 & 5   & 500 & 100 & 100  \\
$b_i$ & 300 & 346 & 373 & 405 & 298 & 313 & 302
\end{tabular}
\end{center}

\noi
We have
\[
\max\left\{\sum_{i\in \{1,2,3,4,5,8,11,12\}}{a_i}\ ,\ \max_{i\in \{6,7,9,10,13,14,15\}}\{b_i\}\right\}=365\ ,
\]
which 
yields a better bound than we got for the case of Hamming distance $12$. Again, the concrete 
decompositions can be found at~\cite{web}.
\pe
The proof of Lemma~\ref{hl} also establishes Proposition~\ref{hp}, and hence 
Theorem~\ref{hr}.

\section{Concluding Remarks}

We have shown that the first proposed two-party pseudo-telepathy game, due 
to Brassard, Cleve, and Tapp~\cite{BCT99}, cannot be won classically for 
the parameter $n=4$ (which is hence the smallest parameter with this 
property). Two players on the other hand sharing four 
maximally entangled quantum bits can use these to completely avoid 
the necessary 
communication and win the game. A number of repeated executions of this game 
can, provided that Alice and Bob never fail, be seen as a convincing demonstration 
of the existence of quantum entanglement.


\begin{thebibliography}{10}








\bibitem{BCT99}
G.~Brassard, R.~Cleve and  A.~Tapp, 
``The cost of exactly simulating quantum entanglement 
with classical  communication'', 
{\em Physical Review Letter}, Vol.\ 83, No.\ 9,  pp.\ 1874--1878, 1999. 

\cancel{
\bibitem{BDHT99}
H. Buhrman, W. van Dam, P. H{\o}yer and A. Tapp, 
``Multiparty Quantum Communication Complexity'', 
{\em Physical Review A}, vol. 60(4), October 1999, pp.\ 2737--2741. 
[\textbf{Viktor$>$ The item is referenced nowhere!}]
}

\cancel{
\bibitem{Ga2001}
V. Galliard,
``Classical pseudo-telepathy and coloring graphs'',
Diploma thesis, ETH Zurich, February 2001.
}

\bibitem{gw02}
V.\ Galliard and S.\ Wolf, ``Pseudo-telepathy, entanglement, 
and graph colorings'', in {\em Proceedings of ISIT 2002}, 2002.

\bibitem{EKR61}
P. Erd\"os, C. Ko, and R. Rado,
``Intersection theorem for systems of finite sets'',
{\em Quarterly journal of mathematics}, Vol.\ 2, No.\ 12, 
pp.\ 313--320, 1961. 

\bibitem{Wi84}
R.~M.~Wilson,
``The exact bound in the Erd\"os-Ko-Rado theorem'',
{\em Combinatorica}, Vol.\ 4,  pp.\ 247--260, 1984.

\cancel{
\bibitem{GHZ89}
Greenberger, D. M., Horne, M. A. and Zeilinger,
``Bell's Theorem, Quantum Theory, and Conceptions of the Universe'',
1989, edited by M. Kafatos (Kluwer, Dordrecht), 74.

\bibitem{GHSZ90}
Greenberger, Horne, Shimony and Zeilinger, 
``Bell's Theorem Without Inequalities'',
{\em American Journal of Physics}, Vol 58, pp.113-122 (1990)

\bibitem{Mermin90}
Mermin, N. D.,
``What's Wrong with These Elements of Reality?'',
{\em Physics Today}, June 1990, 9.
}

\bibitem{web}
http://math.galliard.ch/Quantum/pseudo-telepathy/data03/ ptp\_cliques\_twg03.txt. 
In ptp\_cliques\_twg03\_extract.txt, initial vertices and sizes of cliques are given.


\end{thebibliography}
\end{document}